\documentstyle[prl,aps,epsf]{revtex}

\title{Non-abelian statistics of half-quantum vortices in
$p$-wave superconductors}

\author{D. A. Ivanov}

\address{Institut f\"ur Theoretische Physik,
ETH-H\"onggerberg, CH-8093 Z\"urich, Switzerland}

\date{May 11, 2000}

\begin{document}

\maketitle

\begin{abstract}
Excitation spectrum of a half-quantum vortex in a $p$-wave superconductor 
contains a zero-energy Majorana fermion. This results in a degeneracy
of the ground state of the system of several vortices.
From the properties of the solutions
to Bogoliubov-de-Gennes equations in the vortex core we derive
the non-abelian statistics of vortices identical to that for 
the Moore-Read (Pfaffian) quantum Hall state.
\end{abstract}

\bigskip

Certain types of superconductors with triplet pairing allow
half-quantum vortices \cite{Volovik-halfquantum}.
Such vortices appear if the multi-component order parameter has extra
degrees of freedom besides the overall phase, and the vortex
involves both a rotation of the phase by $\pi$ and a rotation of the
``direction'' of the order parameter by $\pi$, so that the
order parameter maps to itself on going around the vortex.
The magnetic flux through such a vortex is one half of the
superconducting flux quantum $\Phi_0$.

Another signature of this unusual flux quantization is
a Majorana fermion level at zero energy inside the vortex core
\cite{Read-Green}. This energy level has a topological nature
\cite{Volovik-zeromodes} and from the continuity considerations
must be stable to any local perturbations.
In terms of the energy levels, the Majorana fermions in vortex cores
imply a $2^n$-fold degeneracy of the ground state of a system with
$2n$ isolated vortices. If we let vortices adiabatically move around 
each other, this motion may result in a unitary transformation in the
space of ground states (non-abelian statistics). We shall see that
it is indeed the case.

The non-abelian statistics for half-quantum vortices has been originally
derived for the Pfaffian quantum Hall state proposed by Moore and Read
\cite{Moore-Read}. The Pfaffian state is of Laughlin type and may be
possibly realized for filling fractions with even denominator. The
excitations in the Pfaffian state are half-quantum vortices, and their
non-abelian statistics has been derived in the field-theoretical
framework \cite{Nayak-Wilczek,Fradkin,Read-Rezayi,Gurarie-Nayak}.

On the other hand, recently Read and Green suggested that the Pfaffian
state belongs to the same topological class as the BCS pairing state
and thus the latter must have the same non-abelian statistics 
\cite{Read-Green}. In our paper we verify this directly in the BCS
framework as the property of solutions to Bogoliubov-de-Gennes equations.
Our derivation provides an alternative (and possibly more transparent)
point of view on the non-abelian statistics of half-quantum vortices
as well as an additional verification of topological equivalence between
Pfaffian and BCS states.

\bigskip

Let us begin our discussion with reviewing the properties of a half-quantum
vortex. To be specific, we consider a chiral two-dimensional superconductor
with the order parameter of $A$ phase of $^3$He. The order parameter is
characterized by the direction ${\bf \hat d}$ of the spin triplet
(the projection on which of the spin of the Cooper pair is zero)
and by the overall phase $\varphi$. The wave function of the condensate
is
\begin{equation}
\Psi_\pm = e^{i \varphi}
\bigg[ 
d_x \Big( \left|\uparrow\uparrow\right\rangle + 
\left|\downarrow\downarrow\right\rangle \Big) +
i d_y \Big( \left|\uparrow\uparrow\right\rangle - 
\left|\downarrow\downarrow\right\rangle \Big) +
d_z \Big( \left|\uparrow\downarrow\right\rangle + 
\left|\downarrow\uparrow\right\rangle \Big) 
\bigg]
(k_x \pm i k_y).
\label{A-phase}
\end{equation}
The $\pm$ signs denote the two possible chiralities of the condensate.
The chirality breaks the time-reversal symmetry and means a non-zero
angular momentum of the Cooper pairs. In a physical chiral superconductor
there must exist domain walls separating domains of opposite chirality.
Experimentally, domain walls may possibly be expelled from the sample
by an external field which makes one of the chiralities energetically
favorable. In our discussion we do not consider interaction of vortices
with domain walls, but assume that the chirality is fixed over the region
where the vortex braiding occurs (and takes positive sign in 
eq.(\ref{A-phase})).

For the half-quantum vortex to exist, the vector ${\bf \hat d}$
must be able to rotate (either in a plane or in all three dimensions).
An important observation is that the order parameter maps to itself
under simultaneous change of sign of the vector ${\bf \hat d}$
and shift of the phase $\varphi$
by $\pi$: $(\varphi,{\bf \hat d}) \mapsto (\varphi+\pi,-{\bf \hat d})$. 
The half-quantum
vortex then combines rotations of the vector 
${\bf \hat d}$ by $\pi$ and of the phase $\varphi$ by $\pi$
on going around the vortex core (Fig.~1).
This vortex is topologically stable, i.e. it cannot be removed
by a continuous (homotopic) deformation of the order parameter.

Without loss of generality, we consider the vector ${\bf \hat d}$
rotating in the $x$-$y$ plane. The direction of the rotation of the
phase $\varphi$ may either coincide or be opposite to the chirality
of the condensate, which defines either a positive ($\Phi=1/2$)
or a negative ($\Phi=-1/2$) vortex respectively.

There are also two possible directions of rotating the vector 
${\bf \hat d}$. If the vector ${\bf \hat d}$ is confined to a plane
(i.e. takes values on a one-dimensional circle) by an anisotropy
interaction, this gives two possible winding numbers of the
vector ${\bf \hat d}$ ($m=\pm1/2$). If the vector 
${\bf \hat d}$ is not confined to a plane, but takes values on
a two-dimensional sphere, the two directions of winding vector
${\bf \hat d}$ are topologically equivalent, and there is no
additional quantum number characterizing the vortex.

\begin{figure}
\epsfxsize=2.5in
\centerline{\epsffile{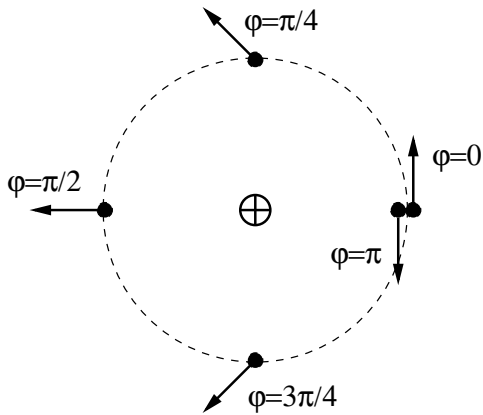}}
\smallskip
\caption{Half-quantum vortex. Arrows denote the direction of vector
${\bf \hat d}$.}
\label{fig1}
\end{figure}

We further restrict our discussion to $\Phi=1/2$ vortices
($\Phi=-1/2$ vortices have a slightly different structure
of the quasiparticle eigenfunctions, but their spectrum
and braiding statistics are the same as for $\Phi=1/2$
vortices). At such a vortex, the condensate wave function
(\ref{A-phase}) takes the form:
\begin{equation}
\Psi(r,\theta)=\Delta(r)\Big[ e^{i\theta}
\left|\uparrow\uparrow\right\rangle +
\left|\downarrow\downarrow\right\rangle \Big]
(k_x+i k_y).
\label{vortex}
\end{equation}
Here $r$ and $\theta$ are the polar coordinates in the vortex,
the windings of the phase $\varphi$ and of the vector ${\bf \hat d}$
have been taken into account.

The Bogoliubov-de-Gennes Hamiltonian in this case decouples into
the two Hamiltonians for spin-up and spin-down electrons. The
spin-down sector contains no vortex and no subgap states.
The Hamiltonian of the spin-up sector is
\begin{equation}
H=\int d^2r \left[
\Psi_\uparrow^\dagger \left({p^2\over 2m} -\varepsilon_F\right)
\Psi_\uparrow + e^{i\theta} \Delta(r) \Psi_\uparrow^\dagger
\left(\nabla_x+i\nabla_y\right)\Psi_\uparrow^\dagger + {\rm h.c.} \right].
\end{equation}
Thus it may be considered as a single-quantum vortex in a superconductor
of spinless fermions.
The quasiparticles do not have a definite spin projection, but are
mixtures of spin-up electrons and spin-down holes:
\begin{equation}
\gamma^\dagger=u\Psi^\dagger_\uparrow + v\Psi_\uparrow.
\end{equation}
Bogoliubov-de-Gennes equations for $(u,v)$ are obtained as
$[H,\gamma^\dagger ]= E\gamma^\dagger$. They are identical to
those for a single-quantum vortex (with the vector ${\bf \hat d}$
constant in space) and were solved by Kopnin and Salomaa in the
context of superfluid $^3$He vortices \cite{Kopnin}. The low-energy
spectrum is
\begin{equation}
E_n=n\omega_0,
\end{equation}
where $\omega_0 \propto \Delta^2/\varepsilon_F$ is the level spacing.
The quantum number $n$ takes integer values (which distinguishes $p$-wave
vortex states from Caroli-de-Gennes-Matricon states in $s$-wave vortices
\cite{deGennes}) and has the meaning of the angular momentum of the
quasiparticle.

In contrast to the single-quantum vortex considered by Kopnin and
Salomaa, in the half-quantum vortex there is an additional relation
between positive- and negative-energy eigenstates, namely
$\gamma^\dagger(E)=\gamma(-E)$. In other words, the solutions with
positive and negative energies are the creation and annihilation
operators for the same fermionic level. Therefore the number of degrees
of freedom in a half-quantum vortex is one half of that in a single-quantum
vortex. The zero-energy level becomes a Majorana fermion:
\begin{equation}
\gamma^\dagger(E=0)=\gamma(E=0).
\end{equation}

It is worth mentioning that this Majorana fermion is stable with respect
to any local perturbation including external potential, electromagnetic
vector potential, local deformations of the order parameter, spin-orbit
interaction and Zeeman splitting (in a single-quantum vortex, only the
first three of those perturbations preserve the zero-energy level
\cite{Ivanov}). We can easily prove it with continuity considerations.
Indeed, suppose that we gradually increase perturbation to the
vortex Hamiltonian (which includes both the spin-up and spin-down sectors).
The levels will shift and mix, but they must do it continuously,
and therefore the number of levels is preserved. Since it is 
half-integer without perturbation, it must remain half-integer for
the perturbed Hamiltonian, i.e. the Majorana fermion survives the
perturbation. This argument is valid as long as the perturbation is
sufficiently small so that the low-lying states remain localized in
the vortex.

Before we turn to discussing the non-abelian statistics of vortices,
let us see how the Majorana fermion $\gamma(E=0)$ transforms under
$U(1)$ gauge transformations. If the overall phase of the superconducting
gap shifts by $\phi$, it is equivalent to rotating electronic creation
and annihilation operators by $\phi/2$: $\Psi_\alpha\mapsto 
e^{i\phi/2} \Psi_\alpha$, $\Psi_\alpha^\dagger \mapsto 
e^{-i\phi/2} \Psi_\alpha^\dagger$. The solution $(u,v)$ transforms 
accordingly: $(u,v)\mapsto (ue^{i\phi/2}, ve^{-i\phi/2})$. The
important consequence of this transformation rule is that under change
of the phase of the order parameter by $2\pi$ the Majorana
fermion in the vortex changes sign: $\gamma\mapsto -\gamma$. This
is an obvious consequence of the fact that the quasiparticle is a linear
combination of fermionic creation and annihilation operators carrying charge
$\pm 1$.

Now consider a system of $2n$ vortices, far from each other
(at distances much larger than $\xi_0\sim v_F/\Delta$). To each
vortex there corresponds one Majorana fermion (further we shall denote
them by $c_i$, $i=1,\dots,2n$) commuting with the Hamiltonian.
They can be combined into $n$ complex fermionic operators and
therefore give rise to the degeneracy of the ground state equal to
$2^n$ (each fermionic level may be either filled or empty). If the
vortices move adiabatically slowly so that we can neglect transitions
between subgap levels, the only possible effect of such vortex motion
is a unitary evolution in the space of ground states.

Let us fix the initial positions of vortices. Consider now a permutation
(braiding) of vortices which returns vortices to their original positions
(possibly in a different order). Such braid operations form a braid
group $B_{2n}$ (multiplication in this group corresponds to the sequential
application of the two braid operations) \cite{braids}. This group
may be described formally in the following way.

Suppose we order all particles along a fixed non-self-intersecting
path. Then braid operations are generated by elementary interchanges
of two neighboring particles. Denote such an elementary operation 
interchanging particles $i$ and $i+1$ by $T_i$ ($i=1,\dots,2n-1$).
Then the group $B_{2n}$ is generated by operators $T_i$ modulo
the following relations (see Fig.~2):
\begin{eqnarray}
T_i T_j &=& T_j T_i, \qquad |i-j|>1, \nonumber\\
T_i T_j T_i &=& T_j T_i T_j, \qquad |i-j|=1.
\label{braid-comm}
\end{eqnarray}

\begin{figure}
\epsfxsize=2.8in
\centerline{\epsffile{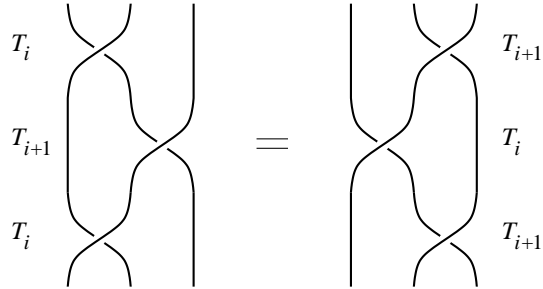}}
\smallskip
\caption{Defining relation for the braid group: $T_i T_{i+1} T_i
= T_{i+1} T_i T_{i+1}$.}
\label{fig2}
\end{figure}

The braiding statistics is defined by the unitary operators in the
space of ground states representing the braid operations from $B_{2n}$.
Here an important reservation has to be made. When a single vortex 
moves along a closed loop, the multi-particle state acquires a phase
proportional to the area inside the loop (every electron inside the loop
effectively moves around the vortex). We shall disregard this effect and,
as a consequence, loose information about the overall phase of the wave
function. In other words, we shall speak about only a {\it projective}
representation of the braid group $B_{2n}$. However, since the
representation is multi-dimensional, the resulting projective
representation is still nontrivial and transforms different states
into each other --- which implies the non-abelian statistics of vortices.

Since the Majorana fermions $c_i$ change sign under a shift of the
superconducting phase by $2\pi$, we introduce {\it cuts} connecting
vortices to the left boundary of the system (Fig.~3).
We take the superconducting phase single-valued away from the cuts
and jumping by $2\pi$ across the cuts. From examining Fig.~3 one
easily obtains that the transformation exchanging the two
vortices $i$ and $i+1$ (with no vortices between them) changes
the phase of the order parameter at one of the vortices by $2\pi$,
which results in the following transformation rule:
\begin{equation}
T_i:\left\{
\begin{array}{cccc}
c_i & \mapsto & c_{i+1} & \\
c_{i+1} & \mapsto & -c_i & \\
c_j & \mapsto & c_j & \mbox{for $j\ne i$ and $j\ne i+1$}
\end{array}
\right.
\label{rep-operators}
\end{equation}
This defines the action of $T_i$ on Majorana fermions. One easily
checks that this action obeys the commutation relations (\ref{braid-comm}).

\begin{figure}
\epsfxsize=3.0in
\centerline{\epsffile{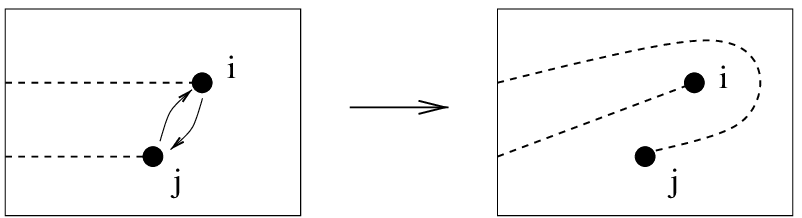}}
\smallskip
\caption{Elementary braid interchange of two vortices.}
\label{fig3}
\end{figure}

Now the action of operators $T_i$ may be extended from {\it operators}
to the Hilbert space. Since the whole Hilbert space can be
constructed from the vacuum state by fermionic creation operators, and
the mapping of the vacuum state by $T_i$ may be
determined uniquely up to a phase factor, the action (\ref{rep-operators})
of $B_{2n}$ on operators uniquely defines a projective representation
of $B_{2n}$ in the space of ground states.

The explicit formulas for this representation may be written in terms
of fermionic operators. Namely, we need to construct operators $\tau(T_i)$
obeying $\tau(T_i) c_j [\tau(T_i)]^{-1} = T_i(c_j)$, where $T_i(c_j)$
is defined by (\ref{rep-operators}). If we normalize the Majorana fermions by
\begin{equation}
\{c_i, c_j\}=2\delta_{ij},
\end{equation}
then the expression for $\tau(T_i)$ is
\begin{equation}
\tau(T_i)=\exp\left({\pi\over4}c_{i+1}c_i\right)
={1\over\sqrt 2}\left(1+c_{i+1} c_i\right)
\label{representation}
\end{equation}
(up to a phase factor).

This formula presents the main result of our calculation. On inspection,
this representation coincides with that described by Nayak and Wilczek
for the statistics of the Pfaffian state \cite{Nayak-Wilczek} (our
Majorana fermions correspond to the operators $\gamma_i$ in section 9
of their paper).

The two simplest examples of the representation (\ref{representation})
are the cases of two and four vortices. These examples were
previously discussed to some extent in the Pfaffian framework in
refs.\cite{Nayak-Wilczek,Fradkin}, and we review them here for
illustration purposes.

In the case of two vortices, the two Majorana fermions may be combined
into a single complex fermion as $\Psi=(c_1+i c_2)/2$, $\Psi^\dagger
=(c_1-i c_2)/2$. The ground state is doubly degenerate, and the only
generator of the braid group $T$ is represented by
\begin{equation}
\tau(T)=\exp\left({\pi\over4} c_2 c_1\right)=\exp\left[i{\pi\over4}
(2\Psi^\dagger\Psi-1)\right]=\exp\left(i{\pi\over4}\sigma_z\right),
\end{equation}
where $\sigma_z$ is a Pauli matrix in the basis ($\left|0\right\rangle$,
$\Psi^\dagger\left|0\right\rangle$).

In the case of four vortices, the four Majorana fermions combine into
two complex fermions $\Psi_1$ and $\Psi_2$ by $\Psi_1=(c_1+i c_2)/2$, 
$\Psi_2 =(c_3+i c_4)/2$ (and similarly for $\Psi_1^\dagger$ and 
$\Psi_2^\dagger$). The ground state has degeneracy four, and the three
generators $T_1$, $T_2$, and $T_3$ of the braid group are represented by
\begin{eqnarray}
 \tau (T_1) &=& \exp\left(i{\pi\over4}\sigma_z^{(1)}\right)
=\pmatrix{e^{-i\pi/4} & & & \cr
          & e^{i\pi/4} & & \cr
          & & e^{-i\pi/4} & \cr
          & & & e^{i\pi/4} }, \nonumber \\
 \tau (T_3) &=& \exp\left(i{\pi\over4}\sigma_z^{(2)}\right)
=\pmatrix{e^{-i\pi/4} & & & \cr
          & e^{-i\pi/4} & & \cr
          & & e^{i\pi/4} & \cr
          & & & e^{i\pi/4} }, \\
 \tau (T_2) &=& \exp\left({\pi\over4} c_3 c_2\right)=
{1\over\sqrt2}(1+c_3 c_2)=
{1\over\sqrt2}\left[1+i(\Psi^\dagger_2+\Psi_2)(\Psi^\dagger_1-\Psi_1)\right]
={1\over\sqrt2}
 \pmatrix{ 1 & 0 & 0 & -i \cr
           0 & 1 & -i&  0 \cr
           0 & -i& 1 &  0 \cr
           -i& 0 & 0 & 1}, \nonumber
\end{eqnarray}
where the matrices are written in the basis
($\left|0\right\rangle$,
$\Psi_1^\dagger\left|0\right\rangle$,
$\Psi_2^\dagger\left|0\right\rangle$,
$\Psi_1^\dagger\Psi_2^\dagger\left|0\right\rangle$).

There are two important properties of the representation 
(\ref{representation}) \cite{Nayak-Wilczek,Fradkin}. The
first one is that $\tau(T)$ are even in fermionic
operators and therefore preserve the parity of the number
of fermions (physically, this simply means that the superconducting 
Hamiltonian creates and destroys electrons only in pairs). Therefore
the representation may be restricted to odd or even sector of the
space of ground states, each of them containing $2^{n-1}$ states
(this degeneracy was also found for the Pfaffian state in refs.
\cite{Nayak-Wilczek,Fradkin,Read-Rezayi}).
Still, in each of these subspaces the representation operators are
non-trivial and non-commuting.

The second property of the representation (\ref{representation})
is that $T_i^4$ is represented by a scalar matrix (projectively 
equivalent to the unity matrix, since we disregard the overall phase).
That is, an elementary interchange of two vortices repeated four times
produces an identity operator (up to an overall phase).

Quite remarkably, our derivation of the non-abelian statistics only
relies on the two facts: first, the flux quantization (half-quantum
for spin-1/2 electrons or, equivalently, single-quantum for spinless
fermions) and, second, that the Majorana fermions carry odd charge
with respect to the vortex gauge field, i.e. they
transform as $c_i\mapsto -c_i$ when the phase of the order parameter
changes by $2\pi$. But these are quantization properties that depend
only on the presence of the Majorana fermion in the vortex spectrum, 
but not on the exact form of the Hamiltonian.
Therefore, if we introduce disorder or other local perturbation
in the BCS Hamiltonian (such as electromagnetic vector potential,
spin-orbit scattering or local deformation of the order parameter),
then not only the Majorana fermions survive, but also the braiding 
statistics (\ref{representation}) remains unchanged (provided the
Majorana fermions stay localized in vortices). Thus we may speak of the
topological stability of the non-abelian statistics (\ref{representation}).

Finally, we mention that the operators $\tau(T_i)$ have also been
discussed in the context of quantum computation as part of a universal
set of operators \cite{Bravyi-Kitaev}. Also, non-abelian anyons
provide a topologically stable realization of unitary operators
for quantum computing \cite{Kitaev}. Thus, should $p$-wave superconductors
with sufficiently large $T_c$ (or, equivalently, large $\omega_0$)
be discovered, they may provide a promising hardware solution for
quantum computation.

The author thanks M.~V.~Feigelman for suggesting this problem
and for many fruitful discussions. Useful discussions
with G.~E.~Volovik, C.~Nayak, I.~Gruzberg, and M.~Zhitomirsky are
gratefully acknowledged. The author thanks Swiss National Foundation for
financial support.

\end{document}